# Investigation of energetics of 1:1 stoichiometric Al/PVP-fuel and nc-CeO$_2$-oxidizer nanocomposite material

S.K. Padhi [a,†], M. Ghanashyam Krishna [a,b]

a. *School of Physics and Advanced Centre of Research in High Energy Materials, University of Hyderabad, Hyderabad 500046, India.* †Email: ***spadhee1@gmail.com***

b. *Centre for Advanced Studies in Electronics Science and Technology, School of Physics, University of Hyderabad, Prof C R Rao Road, Hyderabad 500046, Telangana, India.*

# Table of Contents

# Chapter-5

# Investigation of energetics of 1:1 stoichiometric Al/PVP-fuel and nc-CeO$_2$-oxidizer nanocomposite material

# Chapter-V

# Keywords

**Nanostructured-Al powder, thermo-oxidative reaction, nanostructured ceria, solid-oxidizer, ignition, exotherm, activation energy**

# Abstract

The investigations of the nanostructured-Al (nano-Al) based thermo-oxidative reaction attributes are detailed in this chapter-5. Two products are selected for the thermo-oxidative reaction studies. They are (1) Nano-Al embedded in PVP matrix (will be termed as fuel) and (2) its physically mixed 1:1 stoichiometric nanocomposite processed with the ultrafine nc-ceria oxidizer (also will be termed as nanoenergetic material (NEM)), respectively. The graphical abstract presented below illustrates an overview of the thermo-oxidative reaction of fuel tracked employing thermogravimetric analysis (TGA) investigation. Similar TGA experimentation of the designed NEMs is carried out to gather scientific information validating, whether or not a desired enhanced attribute is achieved. Also, in coincidence with the TGA oxidation event, a synchronous exothermic energy release event in differential scanning calorimetry (DSC) trace is observed. The above mentioned two products and their thermo-oxidative reaction behaviour employing TGA/DTA and TGA/DSC thermal traces is compared and interpreted based on the data available in literature.

# Graphical Abstract

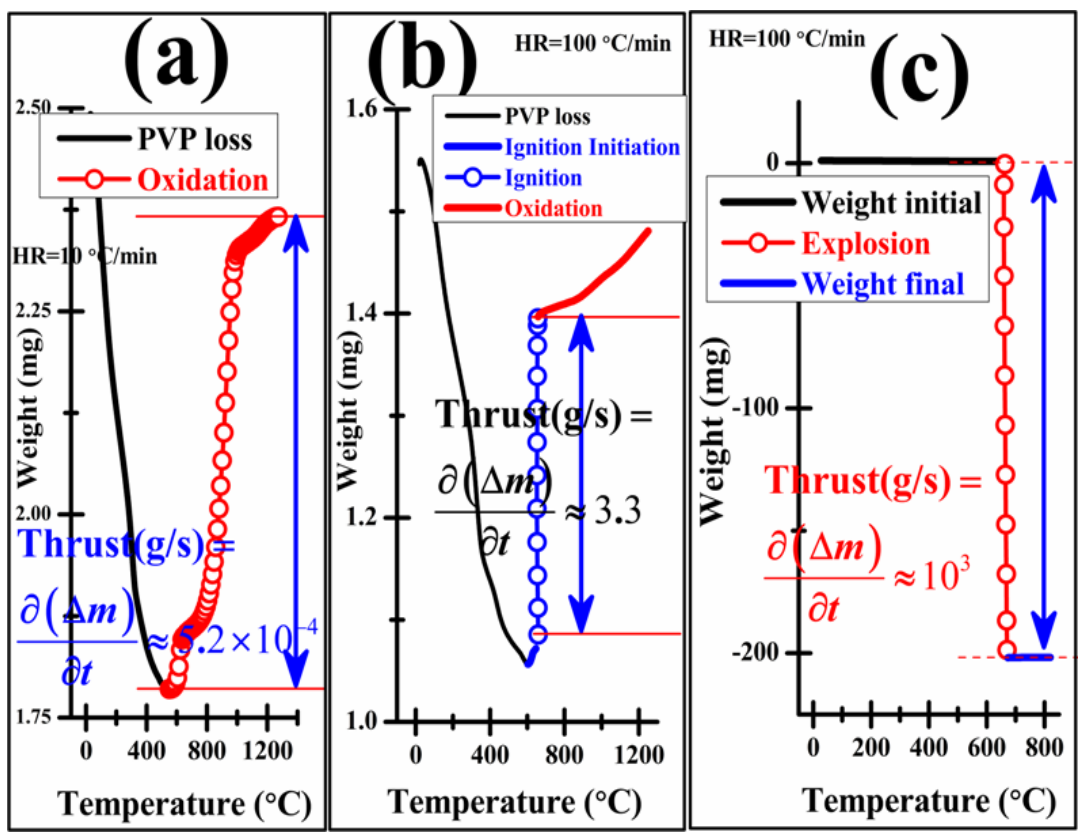

Al/PVP-fuel: (a) Oxidation, (b) Ignition, and (c) Explosion attributes respectively tracked by thermogravimetric thermal analysis (TGA) investigation method.

## 5.1 Introduction

Conventional thermites made of micron-sized metals and the oxide of a less reactive metal as components have low-rate of energy release and long-ignition delay are used in rail road-track welding applications [105]–[107]. The conventional thermite reaction is a diffusion controlled energy release limited by aluminum oxide shell covering the Aluminum (Al) particle [108]–[110]. Pre-stressing of the micron-sized Al core-shell particles to improve reactivity is also attempted. This has shown that flame rate of 68 % is achievable, which is similar to that of for the best case of Al nanoparticles [111]. However, the nanostructured formulation of intimately mixed Al and metal oxide composites dominates the energetic applications [112]–[120]. These composites otherwise termed as metastable intermolecular composites (MICs; also termed as nanoenergetic materials (NEMs)), have enhanced thermo-physical performance as a result of increased interfacial contact area for heterogeneous reaction and reduced diffusion distance. Composite metastability is inertness prior to thermal, laser, or electrical actuation, and also ignition by impact, spark, and frictional force [121]–[127]. For example, when ignited these MICs undergo a self-sustained exothermic reaction to produce almost twice higher volumetric enthalpy (i.e., TNT (conventional explosive) = 7.22 kJcm$^{-3}$, and Al/I$_2$O$_5$ (MIC) = 25.7 kJ cm$^{-3}$) compared to that of the conventional monomolecular high energy explosives [128].

Metallic Al is an ideal, widely utilized, candidate as fuel for example in propellant NEMs, because its high heat of combustion (Enthalpy = 31kJ/g) increases propellant energy density, lowers combustion instability, and facilitates the formation of low molecular weight exhaust gases [129]. It leads to increase in specific impulse of the system at economically lower cost. Also another detailed investigation of a set of NEMs by Zachariah et al. on whether gas phase oxygen generation from oxidizer is a prerequisite for initiation of NEMs reaction had interesting implications [120]. It is found that for Al/Bi$_2$O$_3$ and Al/SnO$_2$ NEMs ignition results below the oxygen release temperature from its corresponding oxidizers, whereas for the second set of NEMs like: Al/Co$_3$O$_4$ results ignition above its oxidizers oxygen release temperature. Further, Al/MoO$_3$, Al/Sb$_2$O$_3$, and Al-WO$_3$ respectively, have oxidizers that did not release any oxygen/ gas. In spite of this, they are seen to ignite showing that oxygen/gas release is a necessary but not sufficient condition which determines the initiation of these NEMs reactions. Therefore, NEMs reaction is the result of direct interfacial contact between Al-fuel and an oxidizer, facilitated by the condensed phase mobility of the reactive species termed as reactive sintering [120], [130]. Many synthetic approaches for NEMs fabrication to develop nano-architecture having Al fuel and oxidizer intimately packed for safer handling include; (1) NEMs composite droplet into core-shell structure [131], (2) filling the oxidizer in protein cages (biothermite) [132], [133], (3) nanowire-based thermite membrane [134], (4) carbon nano-fibers [127], and (5) in the form of both bi-layer/multilayer nanofoil formulations [135]–[139]. These are demonstrated to achieve tunable and efficient NEMs thermo-chemical and energetic properties.

Cerium (IV) oxide (CeO$_2$) is an active candidate widely used in solid oxide fuel-cells and in catalytic converter of toxic species of the automobile exhausts, because of its exceptional reversible reduction-oxidation attribute (2CeO$_2$ ↔ Ce$_2$O$_3$ +1/2 O$_2$) [140]–[146]. Significantly, the performance of low-emission power generation sources such as solid- oxide fuel-cells depend on the ability of nanocrystalline-ceria (nc-ceria) to accept, store, release, and transport oxygen ionic species. Reducing environments lead nc-ceria to form a series of non-stoichiometric oxide phases with Ce$_2$O$_3$ as end reduced product. This end product, in turn, easily can take-up oxygen in oxidizing environment returning back to the full oxidation state. Temperature programmed reduction (TPR) experiments on nc-ceria highlights a four-fold increase in oxygen storage capacity (OSC), and the presence of more reactive surface superoxide (O$^{2-}$) ions [147]–[149]. Theoretical density functional investigations on structure stabilization conducted also reaffirms this experimental evaluated increase in OSC, to fully surface adsorbed supercharged superoxide ions rather than bulk-lattice oxygen species [148]. This supercharging effect is particle size dependent leading to largest OSC for the ultrafine nc-ceria particulate. Thereby a increase in OSC becomes the active source of oxygen species at much lower temperatures (TPR peaks at ≈ 325 °C and 425 °C), in addition to higher bulk and surface lattice oxygen TPR peaks [149]. The unique fascinating properties of nc-ceria like larger OSC and its facile release are the attributes which form the main motivation of the current work to investigate nc-ceria as an oxidizer for NEMs. The synthesized poly (vinylpyrrolidone) (PVP) stabilized nm-Al composite is used as fuel [425]. To the best of the current author's knowledge this is one of the first attempts to employ nc-ceria in NEMs for thermo- physical and energetic enhancements

## 5.2 Materials and Methods

Non-isothermal oxidation at different heating rate runs are acquired by employing TA Instruments STD Q600 dual DSC/TGA (differential scanning calorimetry/thermal gravimetric analysis) instrument. For TGA/DSC, 2.5(±0.2) mg powder samples are taken in alumina cups (90 µL) at a constant flow (100 cm$^3$/min) of nitrogen. The TGA/DSC runs is carried out from ambient temperature to 1573.15 K. TGA and DTA (differential thermal analysis) sensitivity are 0.1µg and 0.001 °C, respectively. DSC calorimetric accuracy is of ±2 % based on the matal standard utilized. Field emission scanning electron microscope (FESEM) [Zeiss-make Ultra 55 model, Everhart-Thornley detector for SE2 and BS electron imaging (20 eV – 30 KeV)] and with attached Oxford Instruments INCA 350 Energy Dispersive X-ray Spectrometer (EDX/EDS) probe are employed for microstructural and chemical investigation (minimum detectable limit for EDX spectroscopy is of about ~0.2 Wt %). Transmission electron microscopy (TEM, FEI

TECNAI G² S-Twin ) in both bright field (BF) and dark field (DF) mode is employed for microstructural detailing at an accelerating voltage of 200 kV.

## 5.3 Results and Discussions

### 5.3.1 NEMs Physically Mixed Reaction Investigation

Physical-mixing is the most straightforward route to develop NEMs in powder formulation [426]–[429]. The reactive components dispersion in hexane (boiling point 68 °C) are probe-sonicated (VCX-750 W ultrasonic processor, 50 % amplitude, 13 mm solid-probe) for different (i.e., 1, 1 h. 30 m, 2 hrs, and 2 hrs 30 m respectively) durations. The NEM powder form is extracted by evaporating the hexane at 80 °C. The physically mixed composite reaction assessment is carried out by TG/DTA, which is to access information on the oxidation initiation temperature, structural phase-change, and energy release attributes, respectively. The NEMs and a conventional energetic material (CEM; nc-ceria oxidizer plus 40 nm-Al from US Nano) TG/DTA data for comparison are shown in fig. 5 1.

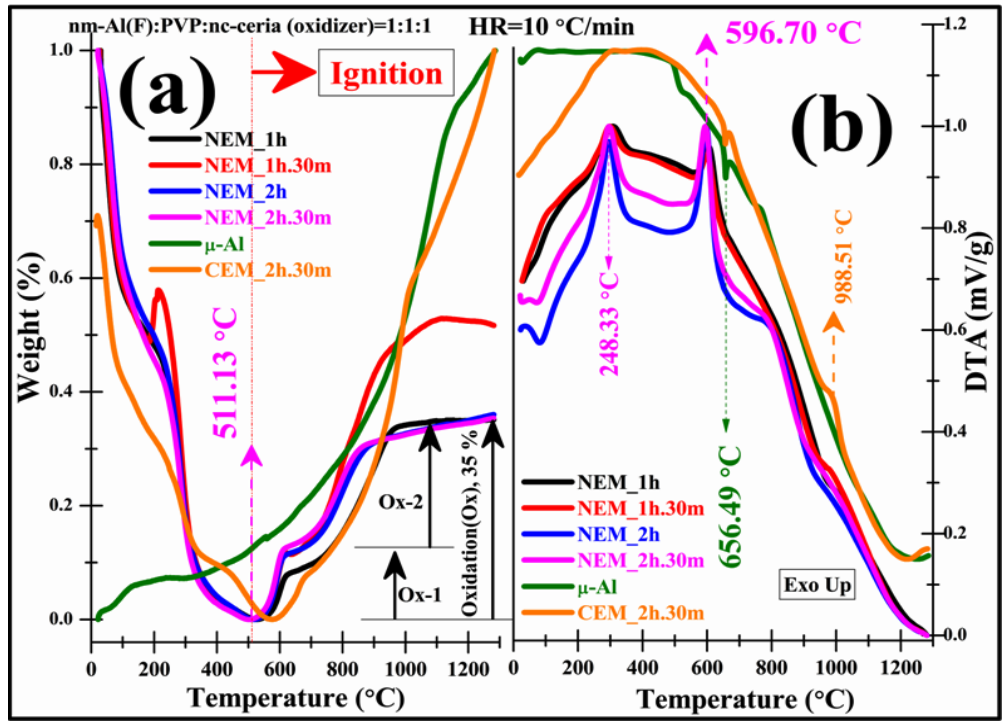

Fig.5 1 NEMs oxidation (a) TG, and (b) corresponding DTA traces respectively

Typically, a batch of approximately 1 g of 1:1 stoichiometric NEMs powder preparation involves physical-mixing of the Al/PVP-fuel with that of the nc-ceria-oxidizer, in hexane as dispersant. The self-heating during sonication is dissipated by maintaining ice-cool water circulation at 20 °C around the sonication vessel. The reaction inferences of this NEM composition based on the TG/DTA plot of fig. 5 1 (a)-(b) are; (1) the oxidation of NEMs initiates at 511.13 °C, and also has a corresponding explicit DTA exotherm at 596.70 °C. (2) The oxidation of NEMs proceeds by two stage-wise weight gain events, i.e., oxidation-1 (Ox-1), and oxidation-2 (Ox-2), is resulting in a total of 35 % of weight gain. Based on the expected Al-fuel fraction, the designed NEMs composition (i.e., nm-Al (Fuel): PVP (stabilizer): nc-ceria (oxidizer) = 1:1:1) should deliver 33 % weight gain, which in this case is close to the 35 % obtained. (3) For NEM ignition at 511.13 °C, the required slope of the Ox-1 or Ox-2 stage must be that of the vertical dotted line drawn (see fig. 5 1 (a)). In other words, initiated oxidation becomes the ignition with an oxidation rate enhancement. (4) Out of the two steps, stepper oxidation like Ox-1 is essential for energy release applications (i.e., have an associated DTA exotherm at 596.70 °C); whereas the gradual slow oxidation event like Ox-2 is not favorable. (5) Once the Ox-1 results in NEM ignition, the exothermic energy release helps in keeping the reaction self-sustained until the total NEM is exhausted. (6) The micron-Al (μ-Al) gradual weight gain is similar to that of the CEM. Both the CEM and μ-Al products indicate Al melting structural phase-transition represented by an endotherm at around 656.49 °C. But (7) in case of our PVP surface stabilized nm-Al product, the melting transition is suppressed and exhibits an oxidation exotherm below it. That is, oxidation is probably by condensed state oxygen delivered from nc-ceria oxidizer. (8) The CEM also has an exotherm after Al melting around 988.51 °C. (9) A low-temperature exotherm observed at 248.33 °C is linked to the weight loss (might be of PVP polymer or nc-ceria grain growth by crystallization or both). The experimental reasoning of having the first exotherm is to be discussed later. (10) Finally, these sets of NEM processed TG/DTA data have a favorable affirmation to use nc-ceria as oxidizer. A minimum 2 hrs physical mixing is adjudged to be the best physical processing duration; to develop 1:1 stoichiometric NEM. The intimately mixed uniformly distributed reactive components inside the PVP are one of the nanocomposites for further evaluation.

### 5.3.2 Nano-Al inherent Surface oxide phase (i.e., in μ-Al, CEM) as inhibitor

The ineffective exothermal energy release of the CEM made out of a similar stoichiometric 1:1 proportion of nc-ceria oxidizer and 40 nm-Al purchased from US nanomaterials (US1050 product ID made by electrical explosion method have density 0.2 gcm$^{-3}$) as observed is further investigated. The electron microscopy imaging techniques of FESEM and TEM are employed. The obtained microstructural data is presented in fig. 5 2. Also, the micron-Al (μ-Al, (9±0.4) μm) powder 1200 °C oxidation product quenched to $LN_2$ temperature employing a vertical furnace microstructural information is included for comparison.

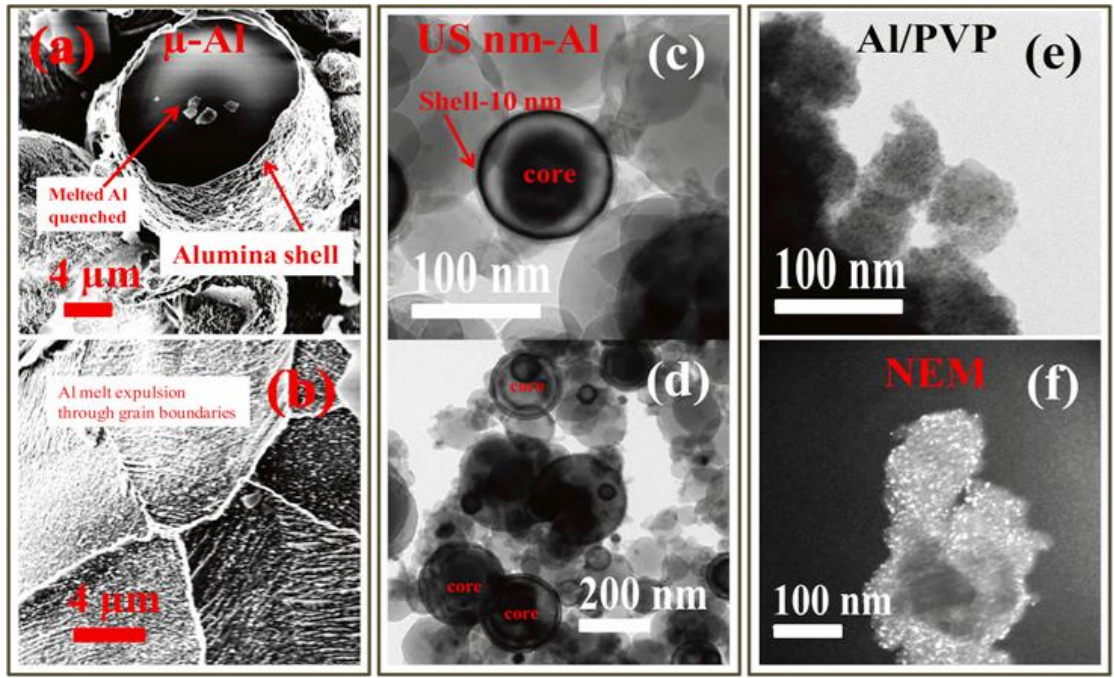

Fig.5 2 Microstructural observations of Al fuel; (a)-(b) µ-Al quenched at 1200 °C, (c)-(d) 40 nm-Al procured from US Nano, and (e)-(f) Al/PVP along with its obtained NEM respectively.

To justify why µ-Al is ineffective in delivering the much needed exotherm for energetic application is linked to the existing surface alumina oxide crust. For elucidation, the µ-Al quenched product powder regions of importance are tracked using FESEM. They are; (1) Al-melt expulsion through particles grain boundaries and isolation of many broken alumina bowls (see fig. 5 2 (a), ruptured alumina shell) with solidified Al-melt is observed. (2) The oxide crust surface shown in fig. 5 2 (b) has clearly defined grain boundaries with well separated highlights of individual grains interior comprising filled nanoscale whiskers, needles, and triangle edged feathers. These local structures grow mostly by outward diffusion of $Al^{3+}$ cations through the Al particle surface structural defect sites and transforms finally into $\alpha$-$Al_2O_3$ phase with oxidation duration and hold time. Tolpygo et al. have examined these solidified Al-melt expulsions and have reported that even after the transformation to α-$Al_2O_3$, these observed structural features are retained for longer duration, depending on the oxidation temperature and surface diffusion processes collectively [430]. Thereby inefficacy to deliver exothermic process out of the μ-Al oxidation progress is mostly linked to the oxide crust diffusion-controlled gradual core oxidation (chemical reaction), and the accompanying structural phase transformation [108], [110]. After Al-core melting, volume expansion outward thrust ruptures the oxide shell (alumina melting=2072 °C) then bringing the already melted Al-core for ignition. Likewise, in the designed CEM, almost similar is the case. That is because of (1) the existence of 10 nm thick alumina oxide shell, and (2) the obtained Al-fuel is of least 0.2 gcm$^{-3}$ densities, thereby no exothermicity in CEM DTA trace is evident. The TEM-BF micrographs of 40 nm-Al US Nano product particles are illustration of core-shell type structures and are shown in fig. 5 2 (c) and (d), respectively. The chemically processed Al/PVP fuel utilized is shown in fig. 5 2 (e), and the corresponding physically mixed NEM developed is shown in fig. 5 2 (f) as TEM-DF mode. The densely packed bright spots representing both of fuel and oxidizer crystals inside the PVP matrix (TEM-DF fig. 5 2 (f)) observed, supports intimate contact between these reacting components and, therefore, lead to energetic efficacy. Lastly, the identification in case of μ-Al powder particles (shown in fig. 5 2 (a)-(b), data not shown) Al-core and alumina oxide shell is carried out employing composition by EDX facility. The average EDX elemental composition from over 10 different random regions on the ruptured crust provides, Al= (54.97±1.2) wt. % and O= (45.03±1.2) wt. % quantitatively alike that of 52.93 and 47.03% expected in $Al_2O_3$. Likewise, the solidified melt region is Al rich, and having (91.71±1.62) wt. % of Al by EDX justify remnant unoxidized mass fraction.

## 5.3.2 Al/PVP Fuel and nc-ceria Oxidizer TG/DTA investigation

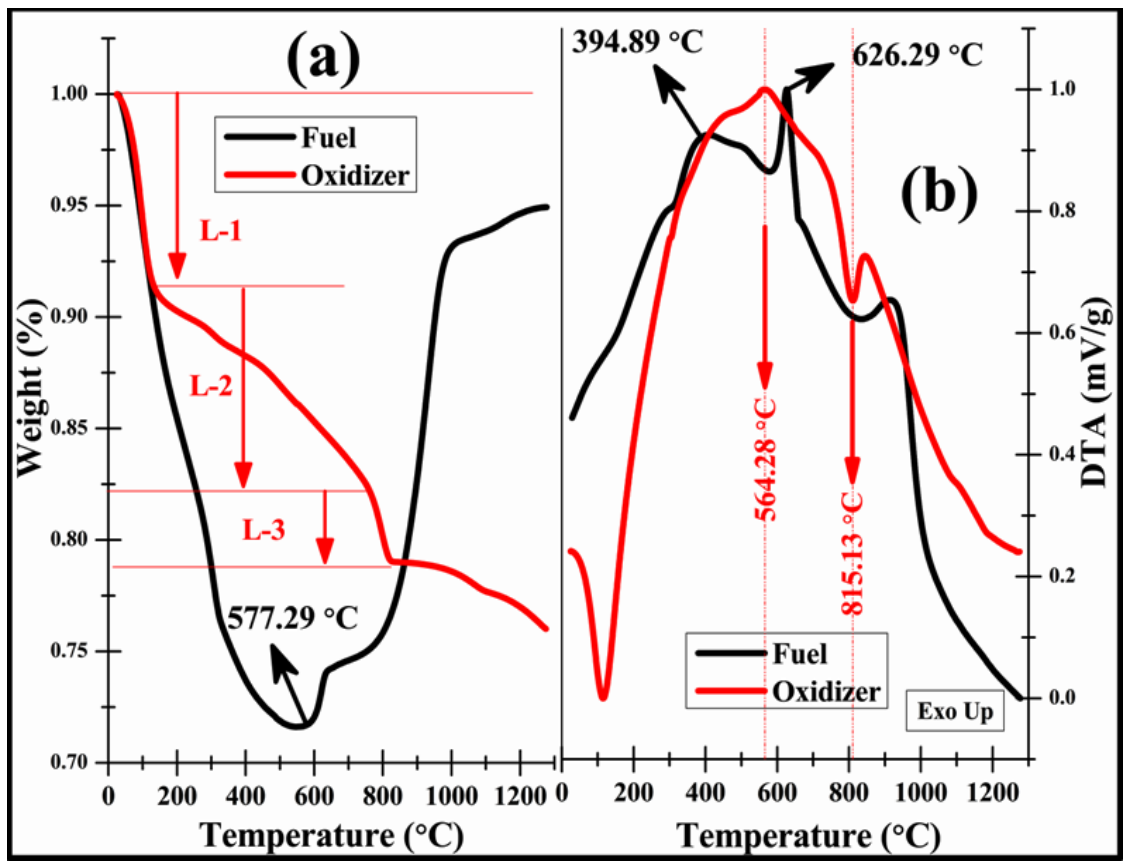

Fig.5 3 TG/DTA analysis of Fuel and Oxidizer at 10 °C/min; (a) TG, and (b) DTA plots respectively.

To have conclusive inference of the observed exotherm at 248.33 °C in the weight loss region of NEMs (see fig. 5 1 (a)), both Al/PVP fuel and nc-ceria oxidizer TG/DTA analysis is carried out. The TG/DTA plots of both are plotted in fig. 5 3 (a)-(b) respectively. The nc-ceria oxidizer has one broad exotherm peaked at 564.28 °C representing gradual weight loss region L-2, whereas the subsequent weight loss region denoted as L-3 is associated with endotherm at 815.13 °C respectively. The weight loss for L-2 is gradual, but L-3 progresses at a faster rate. Based on the previous available literature it is inferred that; (1) first exotherm is probably linked to the gradual release of chemisorbed reactive superoxide ions release at TG L-2 range [431]–[433], whereas (2) endotherm is the lattice oxygen release thereby initiating nc-ceria reduction [434]–[436]. In case of the fuel, the low-temperature range below Al melting has two exotherms at 394.89 and 626.29 °C, respectively. The first exotherm, in this case, is the PVP fraction degradation, whereas the higher exotherm at 626.29 °C is associated with oxidation of embedded Al fraction. A comparison of both HEM and Al/PVP fuel exotherms suggest the nc-ceria oxidizer catalyzes these two processes occurrences to lower temperatures. The PVP degradation lowers to 248.33 °C (oxidizer as catalyst), whereas Al oxidation exotherm is facilitated to 596.70 °C, probably by the release of surface adsorbed superoxide ion release respectively.

### 5.3.3 NEM and Al-PVP Ignition and Energy release investigation

The thermo-chemical reaction properties investigation of both the Al-PVP fuel and NEM are carried out based on TG/DTA and TG/DSC experiments [437]–[444]. In order to present a comparative view both products TG/DSC data recorded at heating rate (HR) at 100 °C/min is plotted in fig. 5 4 (a)-(b). At this one order higher heating rate (i.e. oxidation at 10 °C/min initiates weight gain of 0.52 mg per second for the fuel data shown in fig. 5 3 (a)) than conventional oxidation reaction, the weight gain almost approaches 3.3 g per second. It is about 4 orders higher weight gain than the conventional oxidation, and thereby is termed as ignition. The distinction in these two HR in inducing a sharp exotherm below Al-melting transition (i.e. at 658.46 °C observed for HR=10 °C/min, bulk Al-melting at 660.32 °C) for higher HR is shown in fig. 5 7. Likewise, for HR=100 °C/min designed HEM undergoes initiation of ignition (Ig In) at 369.12 °C and completion at around 397.88 °C. Both exotherms occurs below Al melting; but in the first case, (1) the heterogeneous oxidation (solid and gas as reactants) is achieved in the TGA furnace gaseous environment, whereas in the second (2) intimately mixed condensed nc-ceria donated oxide ions facilities the same. Therefore, the oxidation in presence of nc-ceria becomes homogeneous solid-solid phase reaction. This is induced by condensed solid phase reactive oxygen ions transportation from nc-ceria oxidizer to nc-Al fuel. The nc-ceria induced exotherm in HEM is at 397.88 °C much lower than 654.87 °C, observed for that of fuel.

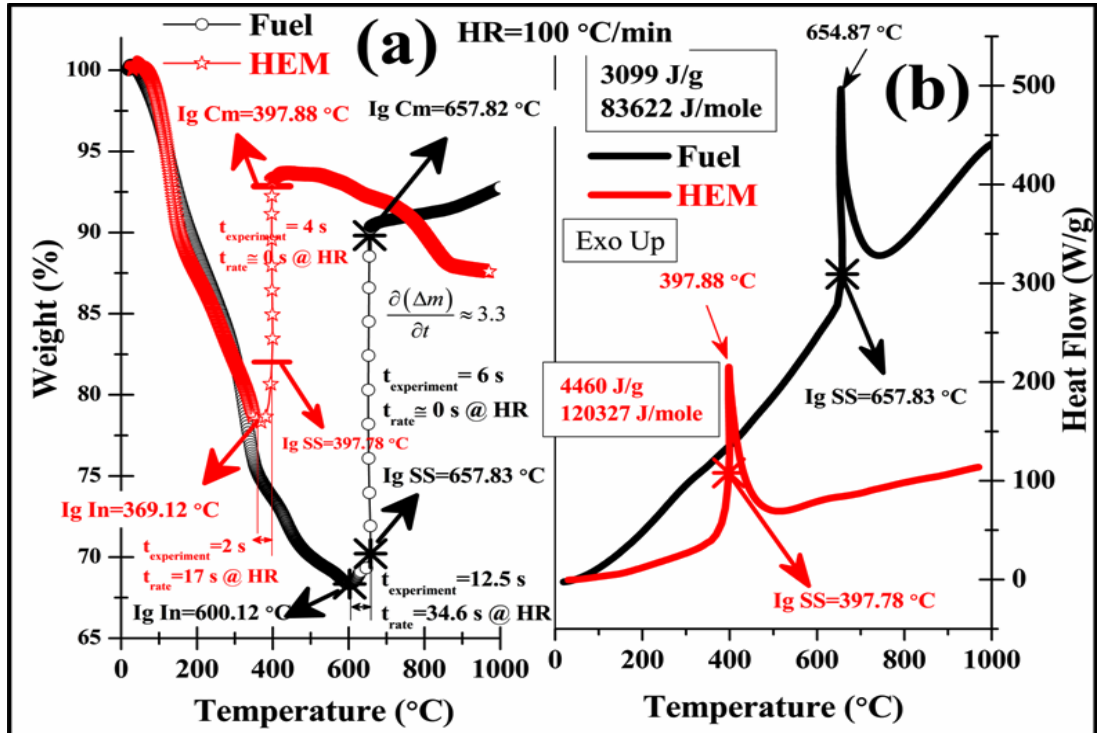

Fig.5 4 TG/DSC analysis at heating rate 100 °C/min; (a) NEM, and (b) Al/PVP respectively.

Also in presence of the nc-ceria solid oxidizer HEM energy release at lower temperature is enhanced by almost about 44 %. A set HR DSC curve of HEM is shown in fig. 5 6 (b), in which the HR generated DSC peak thermal drag (i.e. increase in HR drags the DSC peaks to higher temperature) is used for HEM activation energy extraction. The HEM activation energy for inducing ignition is of 170 kJ/mole which is closer to the activation energy required for the growth of $\gamma$-$Al_2O_3$ phase[445], [446] . The obtained significant energy release enhancement, is believed to be the intimacy in contact achieved between ultrafine 2 nm nc-ceria with that of 3-15 nm Al in PVP matrix. Mostly comparatively larger Al particles surfaces are decorated with many smaller nc-ceria oxidizer all around, as observed from the HR-TEM microstructural detailing (see fig. 5 2 (f) intimacies and density of packing of brighter spots can be seen), supporting the predicted reasoning. In this regard, a table containing ignition characteristics, energy release data obtained and that taken from literature for comparison is tabulated in Table-1.

**Table 5 1** Ignition and Energy release data obtained and existing literature

| NEMs | Ignition Initiation (°C) | Ignition Temperature (°C) | Energy Released (E/T)(in kJ/g) | References |
|---|---|---|---|---|
| Al/CuO | 520 | 540 | 1.8/4.1 | Kim et al [447] |
| Al/NiFe$_2$O$_4$ | 300 | Laser Ignition | 2.9/XX | Shi et al [448] |
| (Al/CuO)/TNT | 242 | 225 | 1.2/XX | Zaky et al [449] |
| Al/CeO$_2$ | 369 | 397 | 4.5/32 | Current study |

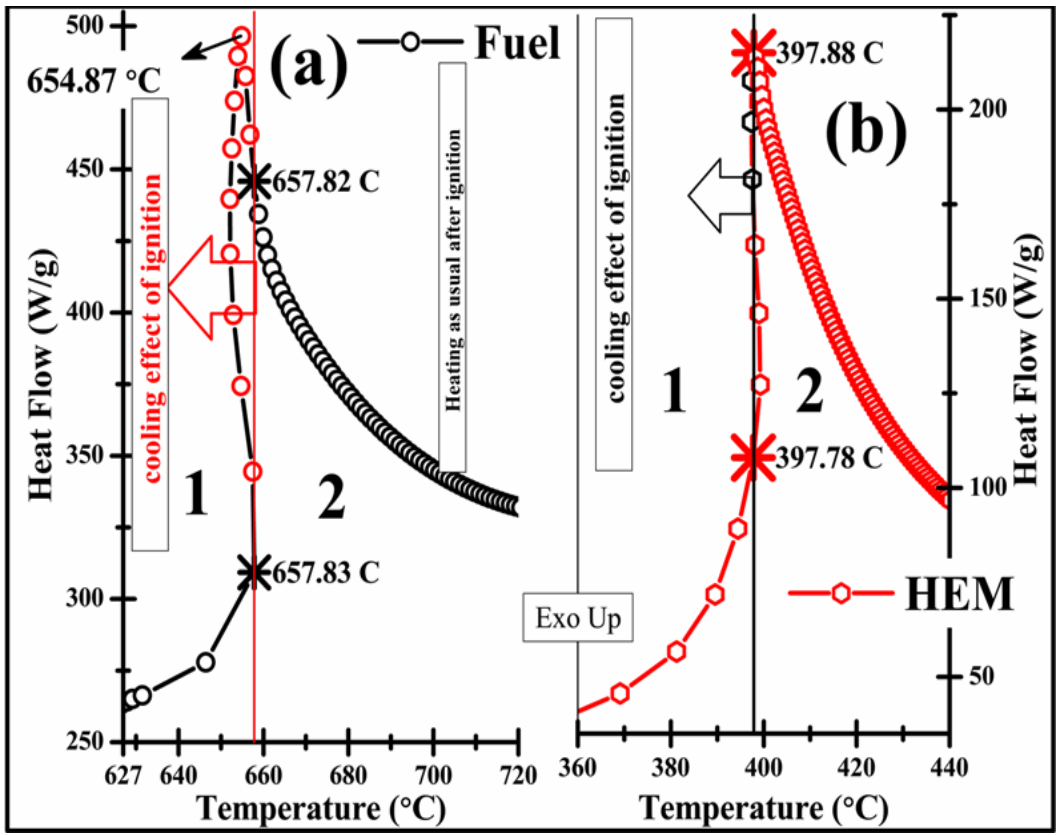

Fig.5 5 DSC data illustrating cooling during ignition both in (a) fuel, (b) HEM respectively.

Besides the tabulated energy release, few other literatures having more or less closest exothermic energy release are; (a) (1) perfluorodecalin coated Al fuel release 4.65 kJ/g , (2) 4.40 kJ/g for toluene coated Al, and (3) 4.20 kJ/g for Al coated with isopropyl alcohol respectively [450], and (b) 4.95 kJ/g for dioctyl sebacate coated Al [451]. Also, the use of carbon nanomaterials (carbon nanotube, expanded graphite, graphene, graphene oxide, and fullerenes) to develop highly energetic compositions is under investigation[452]. The use of carbon nanomaterials involve; (1) reactivity modulation of Al/$WO_3$ EM [453] , (2) 5 wt % graphene oxide in Al/$B_2O_3$ nanocomposite enhances fuel to oxidizer contact resulting enhanced reactivity [454] , (3) 5 wt % carbon black is suitable desensitizer for Al/$MnO_2$ EM [455], (4) tuning gas pressure discharge by adding carbon nanotube in EMs[456], and also (5) tailoring oxidation of Al [446] respectively are achieved. Lastly, it is observed that Al exothermic enthalpy had a strong dependence on CNT content, about 188 kJ/g heat release reported at 20 wt % CNT [446]. The approximately 6-fold higher energy release than 32 kJ/g expected from Al oxidation, is of much interest. It, thus, indicates that the designed NEM from Al-PVP fuel and nc-ceria oxide is quite effective combination in terms of energy release and low temperature ignition, but further enhancement in energy released is possible by designing these two reactive fractions inside graphitic carbon matrix.

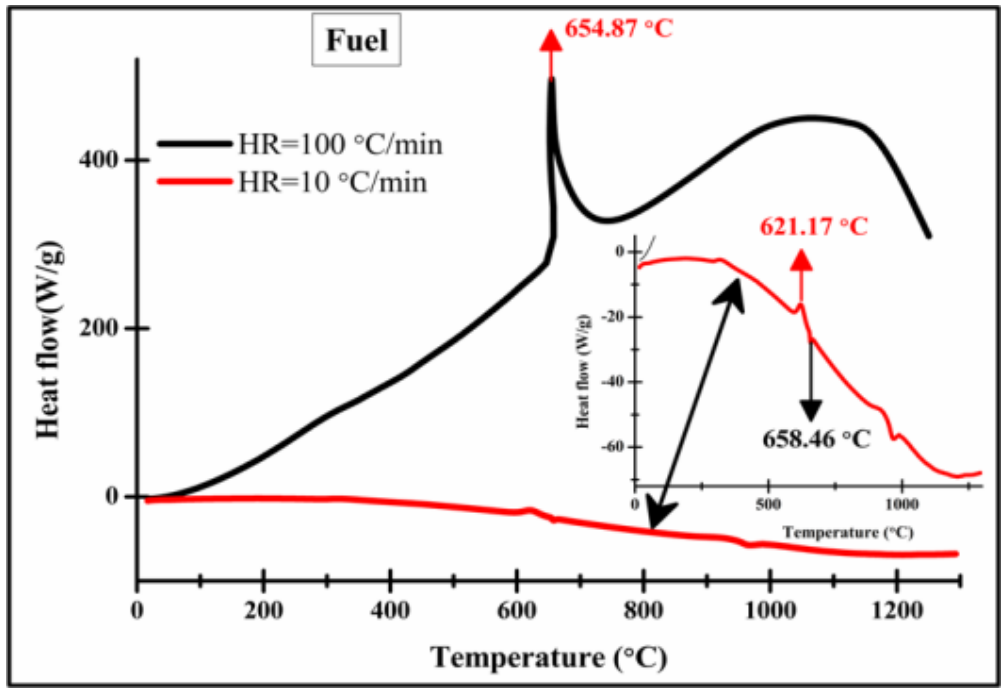

Fig.5 6 DSC data illustrating effect of the HR in igniting the fuel, inset indicates slow HR Al melting transition.

Another significant observation in ignition exotherm is the cooling after ignition, similar to the coined phrase "evaporation causing surface cooling". It is well known that once the ignition is started the subsequent Al oxidation is a self-sustained process. Once self-ignition started the rate of heat generation within the material must over take the rate of heat loss through the material surface[437]. That is, the volumetric heating rate obtained within the material exceeds the experimental heating rate provided, indicating an imbalance resulting few data points of ignition exotherm at lower temperature. These identified data points are in the region-1 of the vertical lines drawn in fig. 5 5 (a)-(b). Likewise, an attempt with higher initial fuel mass of ≥ 10 mg is investigated at HR=100 °C/min indicates the blowing away of the 90 μL alumina sample pan containing HEM. The blowing away happens just after the fuel ignition temperature observed at 657.83 °C, which is evidenced by a sharp weight change of the TGA furnace weight balance (sensitivity=0.1 μg), suggesting occurrence of explosion. 10 mg is the least sample mass required to initiates explosion bringing almost 7 orders higher thrust representing rate of mass change. Thereby, the fuel thrust rate extracted from TG analysis is in the order of; oxidation: ignition: explosion=1:$10^3$:$10^7$ for the employed Al/PVP composite. The Al/PVP explosion representing TG data is shown in fig. 5 7 (a). The TG of explosion is Z- step like appearance no sign of recovery, whereas after oxidation and ignition data recording goes on until the final operating temperature reached.

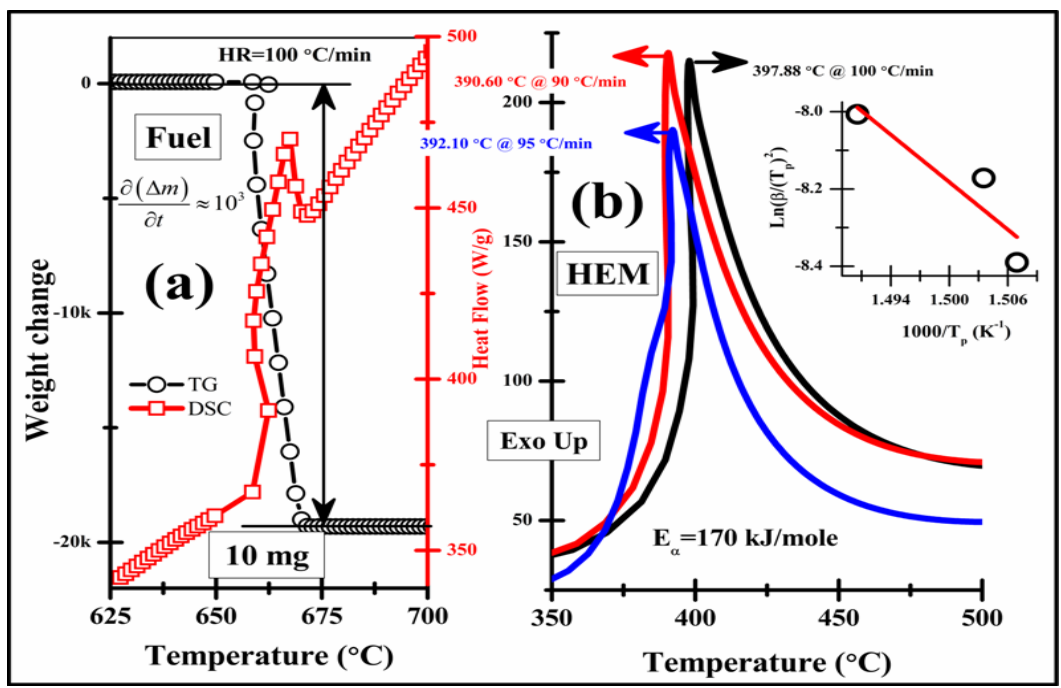

Fig.5 7 DSC data illustrating cooling during ignition both in (a) fuel, (b) HEM respectively.

## 5.4. Conclusions

The evalution of "nc-ceria as an oxidizer" in the stoichiometric nanoenergetic composite formulation has promising implications. The fundamental regenerative adaptability of nc-ceria lattice, whether to act as a source or sink for oxygen release or intake in regards to neighbour is utilized. Specifically, the nc-ceria lattice condensed phase oxygen transfer to nano-Al lattice is established. It leads much lower ignition in the designed HEM at 397 °C than 657 °C, that of the parent Al/PVP fuel. In addition, a jump of 44% in exothermic energy of HEM is achieved. Based on the HR-TEM microstructural detailing the observed intimacy and dense packing of ultrafine 2 nm ceria oxidizer all along the surface of slightly larger 3-15 nm of Al, contained in PVP matrix appears to be the reason for the observed behaviour. The HEM computed activation energy of ignition is 170 kJ/mole, is close to the predicted activation energy required for γ-$Al_2O_3$ phase growth in concurrence to Al oxidation.